\documentclass{aa}
\usepackage{graphicx}
\usepackage{txfonts}
\usepackage{hyperref}
\usepackage{booktabs}
\usepackage{bm}
\usepackage[normalem]{ulem}
\usepackage[dvipsnames]{xcolor}
\usepackage{nicefrac}
\usepackage{mathtools}
\usepackage{hyperref}
\hypersetup{
	colorlinks,
	citecolor=blue,
	linkcolor=blue,
	urlcolor=blue}
\usepackage{amsmath}
\usepackage{yhmath}

\DeclarePairedDelimiter\floor{\lfloor}{\rfloor}

\newcommand{\code}[1]{\texttt{#1}}

\begin{document}
	
	\title{Turbulent dynamo action and its effects on the mixing at the convective boundary of an idealized oxygen-burning shell}
	\titlerunning{Simulations of turbulent dynamo action in a stellar convective shell}

	\author{
		G.~Leidi\inst{1}\and
		R.~Andrassy\inst{1}\and
		J.~Higl\inst{1,2}\and
		P.~V.~F.~Edelmann\inst{3}\and
		F.~K.~R{\"o}pke\inst{1,4}
	}
	\institute{
		Heidelberger Institut f{\"u}r Theoretische Studien,
		Schloss-Wolfsbrunnenweg 35, D-69118 Heidelberg, Germany\\
		\email{giovanni.leidi@h-its.org}
		\and
		High-Performance Computing Center Stuttgart,
		Nobelstraße 19, 70569 Stuttgart, Germany
		\and
		Computer, Computational and Statistical Sciences (CCS) Division and Center for Theoretical
		Astrophysics (CTA), Los Alamos National Laboratory, Los Alamos, NM
		87545, USA
		\and
		Zentrum f\"ur Astronomie der Universit\"at Heidelberg, Institut f\"ur
		Theoretische Astrophysik, Philosophenweg 12, D-69120 Heidelberg, Germany
	}
	
	\date{Received 01 August 2023; accepted 27 September 2023}
	
	\abstract{Convection is one of the most important mixing processes in stellar interiors. Hydrodynamic mass entrainment can bring fresh fuel from neighboring stable layers into a convection zone, modifying the structure and evolution of the star. Because flows in stellar convection zones are highly turbulent, multidimensional hydrodynamic simulations are fundamental to accurately capture the physics of mixing processes. Under some conditions, strong magnetic fields can be sustained by the action of a turbulent dynamo, adding another layer of complexity and possibly altering the dynamics in the convection zone and at its boundaries. In this study, we used our fully compressible \textsc{Seven-League Hydro} code to run detailed and highly resolved three-dimensional magnetohydrodynamic simulations of turbulent convection, dynamo amplification, and convective boundary mixing in a simplified setup whose stratification is similar to that of an oxygen-burning shell in a star with an initial mass of 25 $M_\odot$. We find that the random stretching of magnetic field lines by fluid motions in the inertial range of the turbulent spectrum (i.e., a small-scale dynamo) naturally amplifies the seed field by several orders of magnitude in a few convective turnover timescales. During the subsequent saturated regime, the magnetic-to-kinetic energy ratio  inside the convective shell reaches values as high as 0.33, and the average magnetic field strength is ${\sim}10^{10}\,\mathrm{G}$. Such strong fields efficiently suppress shear instabilities, which feed the turbulent cascade of kinetic energy, on a wide range of spatial scales. The resulting convective flows are characterized by thread-like structures that extend over a large fraction of the convective shell. The reduced flow speeds and the presence of magnetic fields with strengths up to $60\%$ of the equipartition value at the upper convective boundary diminish the rate of mass  entrainment from the stable layer by ${\approx}\,20\%$ as compared to the purely hydrodynamic case.
	}
	
	\keywords{Stars: interiors -- Convection -- Dynamo -- Magnetohydrodynamics (MHD) -- Turbulence}
	
	\maketitle
	%

	\section{Introduction}\label{sec:introduction}
	
	Convection plays a key role in the evolution of stars. In the deep, optically thick layers, convective flows can efficiently transport energy and angular momentum outward, so they determine both the thermal structure and the rotational profile of stars \citep[see, e.g.,][]{maeder2009,kippenhahn2013}. Furthermore, because the characteristic spatial scales of convection are much larger than the mean free path in the stellar plasma, stellar convection zones are highly turbulent environments, with Reynolds numbers that can be as high as $10^{14}$ \citep{jermyn2022}. Turbulent flows quickly mix chemical elements over the relatively short convective turnover timescale, thus profoundly affecting the nuclear energy generation in burning layers of stars and their evolution. 
	
	Despite the huge imprint of convection on stars, most one-dimensional (1D) stellar evolution codes still rely on simplified parametrizations of the convective energy transport, such as the popular mixing-length theory \citep[MLT;][]{prandtl1925,bohm1958}. On 
	the one hand, these parameterized theories allow 1D models to simulate the evolution of stars over thermal and nuclear timescales, which is still unfeasible in multi-D. On the other hand, the parameters that enter these prescriptions cannot be derived from first principles and need to be calibrated. Usually, their value is tuned so that 1D models can reproduce the global properties of our Sun \citep[see, e.g., ][]{richard1996}, but the universality of this approach has been heavily questioned in the literature \citep{trampedach2014,magic2015,joyce2018,sonoi2019}. Moreover, local theories of convection such as the MLT assume that convective mixing stops at the position of the formal convective boundary. More realistically, convective plumes approach the convective boundary with nonzero velocities and give rise to hydrodynamic processes that can entrain some excess mass and entropy from the neighboring stable layer into the convection zone.  Evidence of extra mixing occurring at stellar convective boundaries has been provided by a number of observations, including eclipsing binaries  \citep{valle2016,claret2016}, old open clusters \citep{aparicio1990}, or asteroseismology  \citep{bossini2015,aerts2021}. The entrainment of fresh fuel into a burning layer can prolong its lifetime, enlarge convective cores in upper main sequence stars, and determine the structure of supernova progenitors in more massive stars \citep[][and references therein]{muller2020}.  In stellar evolution codes, mixing at convective boundaries is crudely modeled by means of additional parametrizations, usually in the form of diffusive over-mixing or convective penetration \citep[][and references therein]{anders2023}. The uncertainties arising from the usage of such simplistic models limit the predictive power of stellar evolution calculations and have far-reaching consequences for supernova explosions, the formation of stellar populations, and galactic chemical evolution. Although several nonlocal theories of convection have been presented in the literature \citep[see, e.g.,][]{xiong1978,kuhfuss1986,canuto1997,li2007,garaud2010,canuto2011}, they have not been extensively used in 1D stellar calculations so far.
	
	To overcome the limitations of stellar-evolution models, several research groups have started focusing their efforts in the past two decades on multi-D hydrodynamic modeling of turbulent convection and mixing at convective boundaries in different classes of stellar objects, including core-convective main sequence stars \citep{gilet2013,horst2020a,higl2021,baraffe2023,herwig2023,andrassy2023}, envelope-convective stars \citep{pratt2016,hotta2017,kapyla2019,blouin2023}, and more massive stars during late burning stages \citep{meakin2007,jones2017,cristini2019,andrassy2020,rizzuti2023}. In this approach, nonlinear hydrodynamic processes are captured self-consistently, which allows parameterized theories of convection and convective boundary mixing to be tested and calibrated for different stellar masses and evolutionary stages.

	One more layer of complexity to the problem of stellar convection is, however, represented by the possible presence of magnetic fields, which have been observed both in low- and high-mass stars \citep[][and references therein]{brun2017,keszthelyi2023}. The coupling between turbulent fluid motions and magnetic fields can give rise to small-scale dynamos \citep[SSDs, ][]{meneguzzi1981,brandenburg2005,schekochihin2007}, which amplify magnetic fields on scales smaller than the forcing scale of turbulence. As observed in numerous simulations of solar convection \citep{vogler2007,graham2010,rempel2014,thaler2015,hotta2016,hotta2017}, SSDs can drastically change the morphology of the convective flows, reduce their speed, and alter the dynamics of the overshoot region at the bottom of the solar convection zone as compared to the purely hydrodynamic case. 
	
	Other than the Sun, effects of magnetohydrodynamic (MHD) processes on the properties of convective flows have also been investigated in cool \citep{browning2008,kapyla2021,bhatia2022} and upper-main-sequence stars \citep{brun2005,featherstone2009,augustson2016}, but very few MHD simulations of late burning stages of massive stars have been run to date \citep{varma2020,cuissa2022,varma2023}. In contrast to the main sequence phase, these late evolutionary stages are characterized by vigorous convective shells that can entrain a substantial amount of mass on relatively short timescales, possibly giving rise to shell mergers \citep{ritter2018,mocak2018,yadav2020,andrassy2020}. If an efficient SSD action takes place inside these shells, it could reduce the mass entrainment rate at the convective boundaries and possibly delay or even inhibit the occurrence of the merger events. Numerical simulations of the dynamical amplification of magnetic fields in these layers then become essential for determining the stratification of the supernova progenitor and the fate of the star. 
	Acquiring more insight into dynamo mechanisms in late burning shells of massive stars is also particularly important because magnetic fields can act as seeds to magneto-rotationally powered supernovae \citep{muller2020a}. 
	
	In this paper, we investigated the effects of a small-scale turbulent dynamo acting in a late stellar convective shell using our fully compressible, MHD, \textsc{Seven-League Hydro} (SLH) code. In particular, we used an idealized setup whose stratification is close to that of an oxygen-burning shell in a massive star \citep{andrassy2022}. In this study, we did not intend to perform realistic simulations of such an evolutionary stage, but rather address the following questions: Can an SSD generate dynamically relevant magnetic fields on the typical timescales set by convective motions in the oxygen shell? What is the topology of such fields? What is the feedback of MHD processes on the convective flows and on the mixing at the convective boundary?
	
	The paper is structured as follows: in Sect.~\ref{sec:Methods}, we give a brief description of the numerical methods used to run the simulations needed for this study. The details of the initial stratification are provided in Sect.~\ref{sec:Setup}. In Sect.~\ref{sec:Results}, we present the numerical results, including the evolution of the small-scale turbulent dynamo and its effects on the boundary mixing. Finally, in Sect.~\ref{sec:Conclusions}, we draw conclusions and summarize the main results.

	\section{Methods}\label{sec:Methods}
	
	\subsection{Equations solved}
	
	We described the physical problem by means of the fully compressible equations of ideal MHD with time-independent gravity,
	\begin{align}
		\label{eq:mass_conservation}
		\frac{\partial \rho}{\partial t} + \nabla \cdot (\rho \bm{V}) &= 0,
		\\
		\label{eq:momentum_conservation}
		\frac{\partial  (\rho \bm{V})}{\partial t} + \nabla \cdot  [ \rho \bm{V}
		\otimes \bm{V} + (p+p_B) \mathbb{I} - \bm{B} \otimes \bm{B}] &= \rho \bm{g},
		\\
		\label{eq:Energy_conservation}
		\frac{\partial (\rho e_\mathrm{tot})}{\partial t} + \nabla \cdot [( \rho e_\mathrm{tot} + p + p_B )
		\bm{V} - \bm{B}(\bm{B} \cdot \bm{V}) ]  &= 0,
		\\
		\label{eq:induction_equation}
		\frac{\partial  \bm{B}}{\partial t} + \nabla \cdot  ( \bm{V}
		\otimes \bm{B} - \bm{B}
		\otimes \bm{V} )  &= \bm{0},
		\\
		\label{eq:tracer_conservation}
		\frac{\partial (\rho \psi)}{\partial t} + \nabla \cdot (\rho \psi\bm{V}) &= 0,
	\end{align}
	where $\rho$ denotes the density, $\bm{V}\,{=}\,(V_x,V_y,V_z)$ the velocity vector, $\bm{B}\,{=}\,(B_x,B_y,B_z)$ the magnetic field\footnote{We use the Lorentz-Heaviside units throughout the paper ($\bm{B}\,{=}\,\bm{b}/\sqrt{4 \pi}$).}, $p_B\,{=}\,|\bm{B}|^2/2$ the magnetic pressure, $\bm{g}\,{=}\,(g_x,g_y,g_z)$ the gravitational acceleration, $e_\mathrm{tot}\,{=}\,e_\mathrm{int}+|\bm{V}|^2/2+|\bm{B}|^2/(2\rho)+\phi$ the total energy per unit mass, $e_\mathrm{int}$ the internal energy per unit mass, $\phi$ the gravitational potential, $\psi$ the mass fraction of a passive tracer, and $\mathbb{I}$ the unit tensor. The system in Eqs.~(\ref{eq:mass_conservation})--(\ref{eq:tracer_conservation}) is closed by an Equation of State (EoS) for the gas pressure $p$, 
	\begin{equation}
		p=p(\rho,e_\mathrm{int}).
	\end{equation}
	In our simulations, we assumed an ideal gas law,
	\begin{equation}
		p(\rho,e_\mathrm{int})=(\gamma-1)\rho e_\mathrm{int},
	\end{equation}
	where $\gamma=5/3$ is the adiabatic index. 
		
	We stress that the absence of the viscous and resistive dissipation terms in Eqs.~(\ref{eq:momentum_conservation})--(\ref{eq:induction_equation}) does not mean that the simulated flows are inviscid and nonresistive. In fact, the numerical methods that we used to solve the equations of ideal MHD (see Sect.~\ref{sec:numerical_methods}) must add a certain amount of numerical dissipation into the system in order to achieve numerical stability. 
	
	\subsection{Numerical methods}\label{sec:numerical_methods}
	
	We solved Eqs.~(\ref{eq:mass_conservation})--(\ref{eq:tracer_conservation}) with the \code{SLH} code, which is suited for simulating low-Mach-number (magneto-)convection and excitation of internal gravity waves (IGWs) in the deep interiors of stars \citep{miczek2015a,edelmann2017a,horst2020a,horst2021a,edelmann2021a,andrassy2022,leidi2022,andrassy2023}. \code{SLH} makes use of a second-order finite-volume discretization and upwinding techniques that require an approximate solution to the Riemann problem at each cell interface. Here, the pair of Riemann states was reconstructed using the Piecewise-Parabolic-Method (PPM) of \cite{colella1984}. Upwind, hyperbolic fluxes were computed at cell interfaces with the low-dissipation version of the HLLD solver (LHLLD) of \cite{minoshima2021}. LHLLD modifies the stabilizing pressure-diffusion term in the  original HLLD solver of \cite{miyoshi2005} to ensure that the magnitude of numerical dissipation (relative to the physical central flux) is independent of the Mach number of the flow, $\mathcal{M}=|\bm{V}|/c$, where $c\,{=}\,(\gamma p / \rho)^{1/2}$ is the sound speed. This correction dramatically reduces the excessive amount of numerical diffusion introduced by shock-capturing methods in simulations of subsonic flows \citep{miczek2015a,leidi2022}. 

	To suppress the development of spurious flows due to grid discretization errors in strongly stratified setups, we used a well-balancing technique \citep[the Deviation method,][]{berberich2019, edelmann2021a}. In this method, the vector of conserved quantities, $\bm{U}\,{=}\,(\rho,\ \rho\bm{V},\ \rho e_\mathrm{tot},\ \bm{B}, \ \rho \psi)$, is split into a time-independent hydrostatic component, $\tilde{\bm{U}}$, and a fully nonlinear perturbation, $\delta \bm{U}$. Eqs.~(\ref{eq:mass_conservation})--(\ref{eq:tracer_conservation}) are then solved by enforcing $\partial \tilde{\bm{U}}/\partial t\,{=}\,\bm{0}$, which is achieved in practice by subtracting the hydrostatic fluxes and source terms from the spatial residuals. Such a measure is necessary because conventional finite volume methods discretize hyperbolic fluxes and gravitational source terms at different locations on the computational grid, so hydrostatic solutions cannot be maintained for long times. If ignored, these discretization errors can dramatically affect the evolution of buoyancy-driven flows and produce grossly inaccurate numerical solutions \citep{edelmann2021a}.
	
	To keep the strength of magnetic monopoles under control, we used a staggered constrained transport (CT) method \citep{evans1988}. Different from the finite volume discretization, here the surface-averaged magnetic field components are evolved at cell interfaces by performing the line integral of the electromotive force along the cell edges. Thanks to this operation, the update on the cell-volume average of $\nabla \cdot \bm{B}$ vanishes to machine precision. In \code{SLH}, the upwind electromotive force at the cell edges is computed according to the CT-Contact scheme of \cite{gardiner2005}.
	
	Finally, both cell-centered and staggered quantities were evolved in time with a semi-discrete scheme based on the method of lines. The resulting system of ordinary differential equations was solved using the explicit strong stability preserving (SSP)  RK2 method of \cite{shu1988}. Further details regarding the implementation of the fully unsplit MHD solver in \code{SLH} can be found in \cite{leidi2022}.

	\section{Setup}\label{sec:Setup}
	
	We used the setup first described in \cite{andrassy2022}, who performed a comparison of five different hydrodynamic codes (\code{SLH}, \code{PPMSTAR}, \code{MUSIC}, \code{FLASH}, and \code{PROMPI}) on a problem involving turbulent convection, convective boundary mixing, and the excitation of IGWs in an overlying stable layer. The thermodynamic conditions of this test setup resemble those found during oxygen shell burning in a star with an initial mass of $25\ \mathrm{M}_\odot$ \citep{jones2017}. However, \cite{andrassy2022} adopted a few simplifications to make the study easily reproducible by other research groups. In particular, the geometry of the shell was plane-parallel, the EoS was that of an ideal gas, neutrino cooling was not included, and detailed nuclear burning was replaced by a time-independent heat source term, whose amplitude was set such that convective motions were driven with root-mean-square velocities characteristic of late evolutionary stages in massive stars $(\mathcal{M}_\mathrm{rms}\,{\approx}\, 0.04)$.
	
\begin{figure}
		\includegraphics[width=0.45\textwidth]{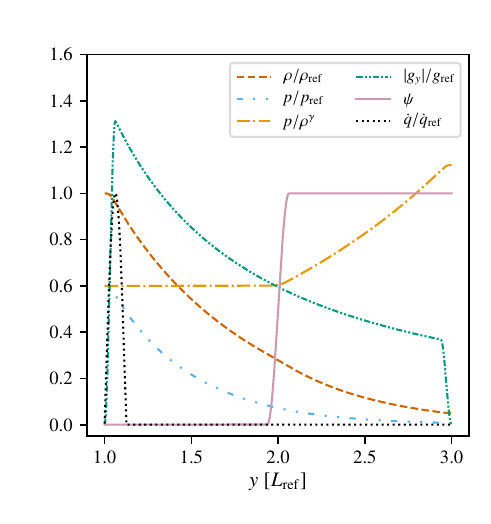}
		\caption{Profiles of density, pressure, pseudo-entropy ($p/\rho^\gamma$), gravity, mass fraction of a passive tracer ($\psi$), and heat source term ($\dot{q}$) as a function of the vertical coordinate $y$ at $t=0\ \mathrm{s}$. Here, $\rho_\mathrm{ref}\,{=}\,1.82\times10^6\ \mathrm{g\ cm^{-3}}$, $p_\mathrm{ref}\,{=}\,4.64\times 10^{23}\ \mathrm{dyne}\ $cm$^{-2}$, $g_\mathrm{ref}\,{=}\,6.37\times10^{8}\ \mathrm{cm\ s^{-2}}$, $\dot{q}_\mathrm{ref}\,{=}\,1.76 \times 10^{20}\ \mathrm{erg}\ \mathrm{cm^{-3}} \  \mathrm{s^{-1}}$, and $L_\mathrm{ref}\,{=}\,4\times10^8$ cm.}
		\label{fig:initial-stratification}
	\end{figure}

	We mapped the initial hydrostatic stratification (see Fig.~\ref{fig:initial-stratification}) on a 3D Cartesian grid with spatial domain $(x,y,z) \in [-L_\mathrm{ref},L_\mathrm{ref}]\times [L_\mathrm{ref},3L_\mathrm{ref}] \times [-L_\mathrm{ref},L_\mathrm{ref}]$, where $L_\mathrm{ref}\,{=}\,4\times10^8 \mathrm{cm}$. We used periodic boundaries in the horizontal $x-$ and $z-$direction. At the top and bottom boundaries of the domain, instead, we adopted impermeable, stress-free boundary conditions for the velocity field,
	\begin{align} 
		\frac{\partial V_x}{\partial y} = \frac{\partial V_z}{\partial y} = V_y = 0, 
	\end{align}
	we forced the magnetic field to be purely horizontal,
	\begin{align} 
		\frac{\partial B_x}{\partial y} = \frac{\partial B_z}{\partial y} = B_y = 0,
	\end{align}
	and for the scalar quantities we assumed
	\begin{align} \label{eq:scalar}
		\frac{\partial \rho}{\partial y} = \frac{\partial p}{\partial y} = \frac{\partial \psi}{\partial y} = 0.
	\end{align}

	The gravitational acceleration, assumed to be time-independent, points downward in the $y$-direction and goes to zero at the vertical boundaries according to Eq. (1) of \cite{andrassy2022}. In that work, such a choice for the gravitational acceleration was made to allow the hydrostatic density and pressure profiles to become constant at the domain boundaries, making the problem consistent with the conditions in Eq.~(\ref{eq:scalar}). Although unrealistic, turning off the gravity at the boundaries does not appreciably alter the stratification of the oxygen shell, which is mostly affected by the aforementioned simplifications, as can be seen in Fig. 1 of \cite{andrassy2022}. For consistency with their model, we decided not to modify the profile of the gravitational acceleration here. 
	
	The stratification is isentropic up to approximately $y\,{=}\,2L_\mathrm{ref}$, and it smoothly turns subadiabatic in the upper half of the domain. Overall, the grid covers $4.35$ pressure scale heights in the vertical direction. To be able to track the time evolution of the mass entrained into the convection zone, we filled the convectively stable layer with a passive tracer at $t\,{=}\,0$ s, whose abundance progressively drops to zero across the convective boundary. Further details regarding the initial stratification and the heat source can be found in \cite{andrassy2022}. 
	
	To start the dynamo action, we planted an initially horizontal magnetic field into the grid, $B_x\,{=}\,10^5\ \mathrm{G}$. The strength of the seed field was chosen such that the Lorentz force exerted on the fluid at early times was weak enough to not affect the development of convection. 
	
	We judged the numerical convergence of the results obtained in this work by running simulations on grids with $128^3$, $256^3$, and $512^3$ cells. To compute meaningful time-averaged quantities and avoid introducing temporal correlations caused by the turbulent nature of the convective flows, all test cases were run until $t_\mathrm{max}\,{=}\,25\tau_\mathrm{conv}$, where $\tau_\mathrm{conv}\,{=}\,63.36\ \mathrm{s}$ is the mean convective turnover timescale in the purely hydrodynamic case, defined according to Eq. (16) of \cite{andrassy2022}. By $t\,{=}\,t_\mathrm{max}$, the growing convective layer is still sufficiently far away from the upper boundary of the spatial domain that the imposed boundary conditions do not appreciably alter the dynamics of the mixing region. Thus, we decided not to extend the simulations beyond $t_\mathrm{max}$. Finally, in order to capture possible differences between the MHD and the purely hydrodynamic case, we ran an additional set of simulations without magnetic fields.
	
	\section{Results}\label{sec:Results}
	
	\subsection{Onset of convection and kinematic stage of the dynamo}

	\begin{figure*}
		\includegraphics[width=1\textwidth]{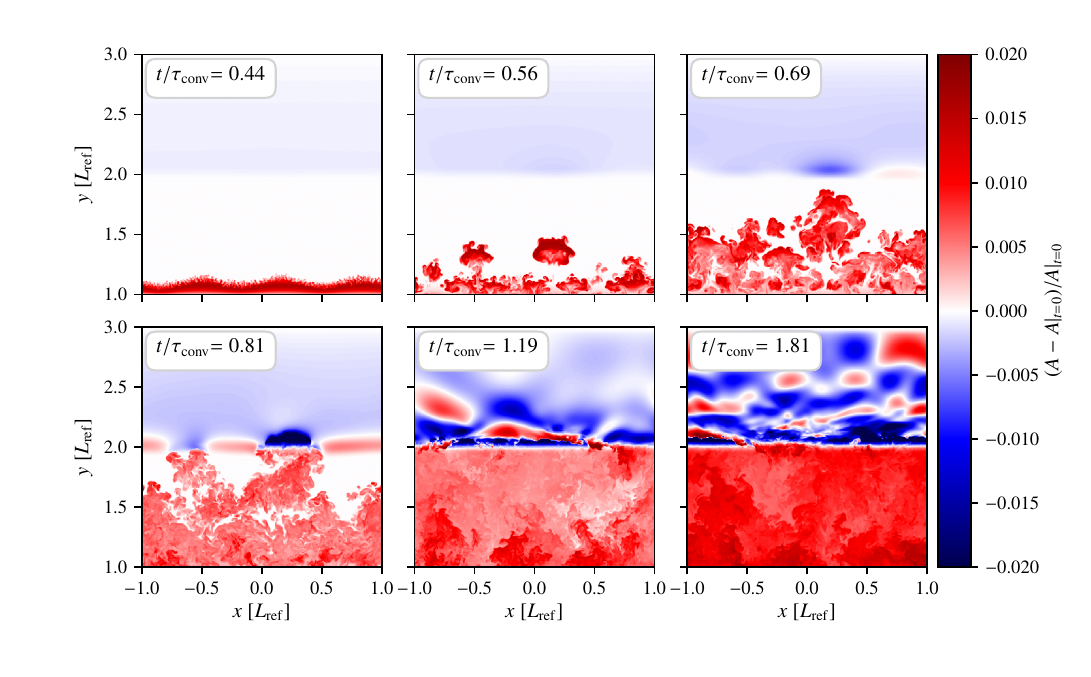}
		\caption{Development of convection in the MHD simulation of the idealized oxygen shell run on a $512^3$ grid. The panels show fluctuations in pseudo-entropy ($A=p/\rho^\gamma$) in the $z\,{=}\,0$ plane at different times, as indicated by the insets. The entropy generated by the heat source  at the base of the box (see Fig.~\ref{fig:initial-stratification}) is mixed throughout the initially adiabatic layer by turbulent convection. This process slowly increases the entropy content of the convection zone in time. The broad stripe of negative entropy fluctuation visible in the upper half of the domain at early times is due  the thermal expansion of the convective layer. The turbulent flows also excite IGWs at the upper convective boundary (lower center panel) which then propagate in the subadiabatic layer.}
		\label{fig:initial-transient}
	\end{figure*}

	To break the initial symmetry, we added a small-amplitude perturbation to the hydrostatic density stratification according to Eq.~(6) of \cite{andrassy2022}. The energy injected by the heat source at the base of the box leads to the development of buoyant parcels of hot fluid that rise in the adiabatic layer (see Fig.~\ref{fig:initial-transient}). As soon as these flows cross the boundary of the subadiabatic layer, the buoyant acceleration changes sign (so it points downward in the $y$-direction) and forces the rising plumes to turn around. The large-scale buoyant fluid elements that are driven by the energy source quickly develop shear instabilities that cascade down to smaller scales, and turbulent convection fully develops by $t\approx\tau_\mathrm{conv}$. 
	
	\begin{figure}
		\includegraphics[width=0.45\textwidth]{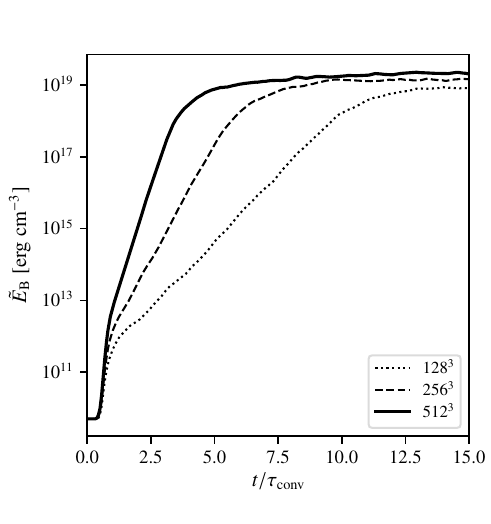}
		\caption{Time evolution of the mean magnetic energy density inside the convection zone for the indicated grid resolutions.}
		\label{fig:growth-rate}
	\end{figure}

	As shown in Fig.~\ref{fig:growth-rate}, the mean magnetic energy density inside the convective shell\footnote{$\langle \cdot \rangle$ is the volume-weighted  spatial average operator.}, 
	\begin{equation}
		\tilde{E}_\mathrm{B} = \frac{1}{2}\langle |\bm{B}|^2 \rangle_\mathrm{conv},
	\end{equation}
	increases exponentially in time. The growth rate of the instability is higher on finer grids, which indicates that the amplification process mostly occurs on intermediate or small spatial scales. It could be an SSD, where the magnetic field is randomly stretched at scales smaller than the forcing scale of turbulence. However, two other processes could contribute to the amplification of small-scale magnetic fields in this setup: turbulent induction, which is the stretching of large-scale magnetic fields by a turbulent, small-scale velocity component \citep{schekochihin2007}, and turbulent cascade of magnetic energy toward smaller scales \citep{graham2010}. In fact, the large-scale fields that are needed to excite the latter two processes, not only are characteristic of large-scale dynamos \citep{brandenburg2009,charbonneau2013}, but they can also be supported by the large-scale velocity structures typical of turbulent convection \citep{kapyla2018}. Moreover, the imposed boundary conditions (see Sect.~\ref{sec:Setup}) allow the integrated horizontal magnetic flux, and consequently the mean horizontal magnetic field, to be preserved in time inside the spatial domain. Therefore, the mean horizontal field takes the value of the chosen seed field, $B_x\,{=}\,10^5\ \mathrm{G}$, and represents a persistent large scale magnetic field component that could, in principle, contribute to the amplification of magnetic energy via turbulent induction.

	\begin{figure*}
		\centering
		\includegraphics[width=0.9\textwidth]{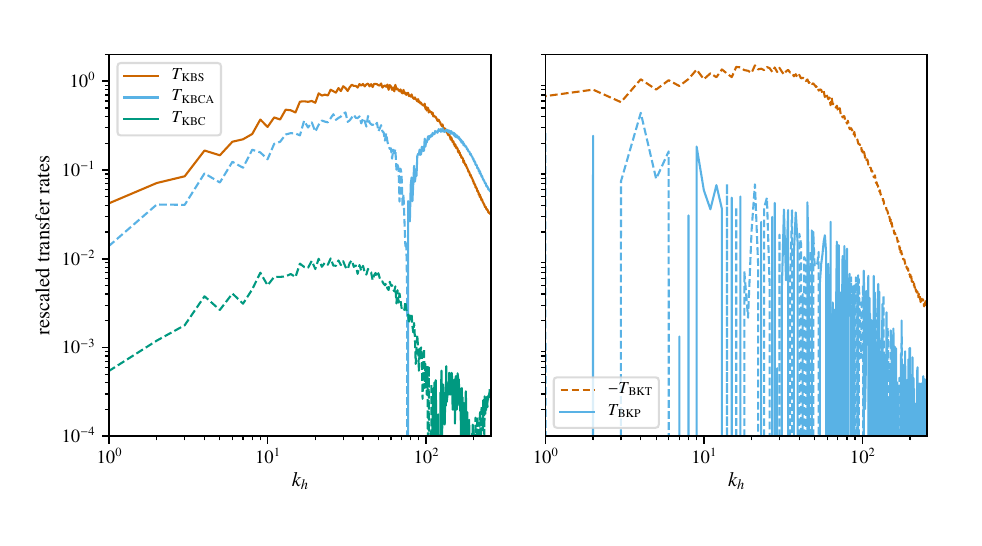}
		\caption{Transfer functions extracted from the convection zone on the $512^3$ grid and averaged over the kinematic stage of the dynamo. The panel on the left shows the transfer rates from the kinetic ($\mathrm{K}$) to the magnetic ($\mathrm{B}$) energy reservoir, whereas the panel on the right shows magnetic-to-kinetic energy transfer rates. The time averaging was performed such that, at each time $t$, all the transfer curves were rescaled by the maximum value of $T_\mathrm{KBS}(k_h)$. The resulting curves were then averaged over the time interval $t \in (1.5\tau_\mathrm{conv},3\tau_\mathrm{conv})$. Dashed lines represent negative transfer rates, while solid lines are used for positive rates.}
		\label{fig:transfer}
	\end{figure*}

	To get a better understanding of the underlying mechanisms that amplify the magnetic energy in these simulations, we computed transfer functions $T_\mathrm{XYZ}(\bm{k})$ in the Fourier space between the kinetic ($\mathrm{K}$) and magnetic ($\mathrm{B}$) energy reservoirs inside the convective layer, following the approach of \cite{graham2010}. In particular, $T_\mathrm{XYZ}(\bm{k})$ represents the energy received (or lost in case of $T_\mathrm{XYZ}(\bm{k})\,{<}\,0$) per unit time and per unit wavenumber  at scale $\bm{k}$ of energy type $\mathrm{Y}$ from all scales of energy type $\mathrm{X}$ via process $\mathrm{Z}$. The transfer of magnetic energy to the $k$-th component of kinetic energy is determined by the net work done on the fluid by the magnetic tension force,
	\begin{equation}
		\begin{split}\label{eq:ta}
			T_\mathrm{BKT}(\bm{k}) = & 
			\frac{1}{2}\hat{\bm{V}}(\bm{k})\cdot[\widehat{\bm{B}\cdot \nabla\bm{B}}]^{*}(\bm{k}) \\
			+  &\frac{1}{2}(\widehat{\rho \bm{V}})(\bm{k})\cdot\Bigg[\widehat{\frac{1}{\rho}\bm{B}\cdot \nabla\bm{B}}\Bigg]^{*}(\bm{k}) + \mathrm{c.c.}, 
		\end{split}
	\end{equation}
	and the magnetic pressure force,
	\begin{equation}
		\begin{split}
			T_\mathrm{BKP}(\bm{k}) = & 
			-\frac{1}{4}\hat{\bm{V}}(\bm{k})\cdot\big[\widehat{\nabla|\bm{B}|^2}\big]^{*}(\bm{k}) \\
			& -\frac{1}{4}(\widehat{\rho \bm{V}})(\bm{k})\cdot\Bigg[\widehat{\frac{1}{\rho}\nabla|\bm{B}|^2}\Bigg]^{*}(\bm{k}) + \mathrm{c.c.},
		\end{split}
	\end{equation}
	where $*$ is the complex conjugate, c.c. is the complex conjugate of the whole expression on the right-hand side, and $\string^$ represents the Fourier projection\footnote{A thorough derivation of the transfer functions computed here can be found in Appendix A.1. of \cite{graham2010}.}. Magnetic energy on scale $\bm{k}$ is produced or removed via stretching of the magnetic field lines, 
	\begin{equation}
		T_\mathrm{KBS}(\bm{k}) = \hat{\bm{B}}(\bm{k})\cdot[\widehat{\bm{B}\cdot \nabla\bm{V}}]^{*}(\bm{k}) + \mathrm{c.c.},
	\end{equation}
	and through compression and advection of the magnetic field, 
	\begin{equation}\label{eq:tb}
		\begin{split}
			T_\mathrm{KBCA}(\bm{k})  = & -\hat{\bm{B}}(\bm{k})\cdot[\widehat{\bm{B}\nabla\cdot\bm{V}}]^*(\bm{k}) \\
			& -\hat{\bm{B}}(\bm{k})\cdot[\widehat{\bm{V}\cdot\nabla\bm{B}}]^*(\bm{k})  + \mathrm{c.c.}
		\end{split}
	\end{equation}
	Because here we solved the fully compressible MHD equations, $ T_\mathrm{KBCA}(\bm{k})$ includes both the transport of energy within the magnetic energy reservoir and the generation of magnetic energy through fluid compression. These two processes, however, cannot be decoupled \citep{rempel2014}, which can be seen by expanding the advective flux of magnetic energy as
	\begin{equation}
		- \nabla \cdot \Bigg( \bm{V}\frac{|\bm{B}|^2}{2} \Bigg)(\bm{k}) = - \hat{\bm{B}}(\bm{k}) \cdot \widehat{\Bigg[ (\bm{V}\cdot\nabla)\bm{B} + \frac{\bm{B}}{2}\nabla \cdot \bm{V}\Bigg]^*}(\bm{k}) + \mathrm{c.c.}
	\end{equation}
	To simplify the calculations, instead of computing 3D Fourier projections in Eqs.~(\ref{eq:ta})-(\ref{eq:tb}), we averaged transfer functions $T_\mathrm{XYZ}(k_h,y_j)$ obtained at each horizontal plane $y_j$ inside the convection zone,
	\begin{equation}
		T_\mathrm{XYZ}(k_h) = \langle T_\mathrm{XYZ}(k_h,y_j)\rangle_{y_j \in (L_\mathrm{ref},2L_\mathrm{ref})}.
	\end{equation}
	Here, $k_h$ is the horizontal wavenumber $k_h=\sqrt{k_x^2+k_z^2}$, where
	\begin{align}
		k_x &= 
		\begin{cases}
			m, & 0 \le m \le \floor*{\frac{N_x - 1}{2}},  \\
			-N_x + m, & \floor*{\frac{N_x - 1}{2}} < m  < N_x, 
		\end{cases} \\
		k_z &= 
		\begin{cases}
			n, & 0 \le n \le \floor*{\frac{N_z - 1}{2}},  \\
			-N_z + n, & \floor*{\frac{N_z - 1}{2}} < n  < N_z, 
		\end{cases} 
	\end{align}
	$\floor*{.}$ is the floor function, and $N_x$ and $N_z$ are the number of cells in the $x-$ and $z-$direction, respectively. 
	
	Figure~\ref{fig:transfer} shows results from the transfer analysis performed on the grid with $512^3$ cells. Stretching of the magnetic field lines contributes most of the magnetic energy generation at spatial wavenumbers close to $k_h\,{=}\,50$. In these simulations, the typical velocities in the convection zone are considerably subsonic ($\mathcal{M}_\mathrm{rms}\,{\approx}\,0.04$), so fluid compression due to the ram pressure of the convective flows ($p_\mathrm{ram}\,{\sim}\,\mathcal{M}^2$) has a negligible contribution to the generation of magnetic energy (see $T_\mathrm{KBC}(k_h) = -\hat{\bm{B}}(k_h)\cdot[\widehat{\bm{B}\nabla\cdot\bm{V}}]^*(k_h)$ + c.c. in Fig.~\ref{fig:transfer}). Therefore, $T_\mathrm{KBCA}(k_h)$ measures  mostly the advective transport of magnetic energy to scale $k_h$ from all scales of the magnetic field, similar to the case of incompressible MHD,
	\begin{equation}
		T_\mathrm{KBCA}(k_h)  \approx -\hat{\bm{B}}(k_h)\cdot[\widehat{\bm{V}\cdot\nabla\bm{B}}]^*(k_h)  + \mathrm{c.c.}
	\end{equation}
	We observe that the magnetic cascade mainly removes magnetic energy from large scales, where $T_\mathrm{KBCA}\,{<}\,0$, and redistributes it at scales with $k_h\,{>}\,75$, where $T_\mathrm{KBCA}\,{>}\,0$. This process dominates the generation of magnetic energy over stretching only at $k_h\,{>}\,130$, which corresponds to a spatial scale of $3.9$ times the width of the grid cells, $\Delta x$. Work done by fluid motions against the magnetic tension force ($-T_\mathrm{BKT}$) most efficiently transforms kinetic energy at $k_h\,{\approx}\,30$ into magnetic energy. Work done by the magnetic pressure force on the fluid is negligible everywhere except on very large scales. These results allow us to find the range of wavenumbers where the magnetic field is most efficiently stretched by fluid motions. As pointed out by \cite{graham2010}, the spatial wavenumber at which magnetic field is generated ($\bm{q}$), the one at which it is stretched ($\bm{k}$), and the one at which the flow works against magnetic tension ($\bm{p}$) form a triadic relation,
	\begin{equation}
		\bm{k} = \bm{q} - \bm{p}. 
	\end{equation}
	By considering the most extreme cases in which $\bm{q}$ and $\bm{p}$ have the same or the opposite orientation, we estimate that the magnetic field lines are most efficiently stretched at $20\,{\lesssim}\,k_h\,{ \lesssim}\,80$. As we show in Sect.~\ref{sec:nonlinear}, this interval lies at the bottom of the inertial range of the turbulent kinetic energy spectrum. Thus, the amplification of magnetic energy is mostly caused by the action of a small-scale turbulent dynamo, with a minor contribution from the turbulent cascade close to the grid scale.

\begin{figure}
		\includegraphics[width=0.45\textwidth]{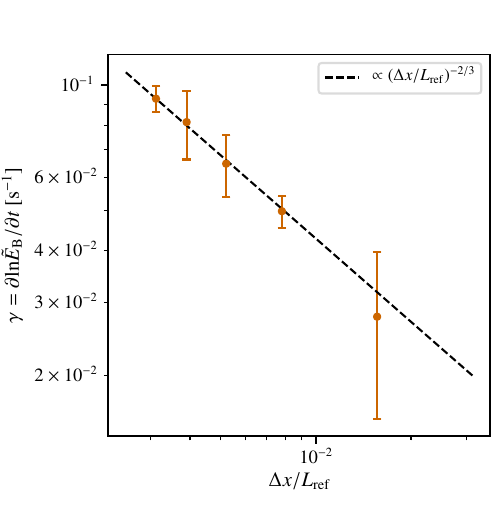}
		\caption{Growth rate of the mean magnetic energy inside the convection zone (averaged over the kinematic phase of the dynamo) as a function of the grid spacing, $\Delta x$. For this analysis, we also simulated the kinematic phase of the dynamo on grids with $384^3$ and $640^3$ cells ($\Delta x / L_\mathrm{ref}\,{=}\,5.2 \times 10^{-3}$, and $\Delta x / L_\mathrm{ref}\,{=}\,3.1 \times 10^{-3}$, respectively). Error bars represent three standard deviations, computed over the time series. The expected theoretical scaling for SSD amplification $(\gamma\propto\mathrm{Re}^{1/2}\propto\Delta x^{-2/3})$ is represented by the black dashed line. }
		\label{fig:gamma}
	\end{figure}

	Further evidence of small-scale turbulent dynamo action can be provided by checking the scaling of the growth rate of the magnetic energy, $\gamma\,{=}\, \partial \mathrm{ln} \tilde{E}_\mathrm{B} / \partial t$, with the grid resolution. Because in this work we used the Implicit Large Eddy Simulation (ILES) method, the magnetic Prandtl number ($\mathrm{Pr}_\mathrm{m}\,{=}\,\nu/\eta$) is likely to be close to or larger than unity \citep{vogler2007,rempel2014}. In this regime of Prandtl numbers, an SSD can only be started if the fluid Reynolds number, $\mathrm{Re}\,{=}\,V_\mathrm{rms} L_\mathrm{ref}/\nu$, is larger than a critical threshold. The growth rate of the magnetic energy in an unstable SSD should then scale as $\mathrm{Re}^{1/2}$ \citep{kazantsev1968,schekochihin2004}. Although the effective value of the kinematic viscosity ($\nu$) and resistivity ($\eta$) coefficients
	are determined by the underlying numerical methods used to solve the MHD equations, in the ILES approach the fluid Reynolds number should depend on the spatial resolution as $\Delta x^{-4/3}$  \citep{cristini2019}, which leads to $\gamma \propto \mathrm{Re}^{1/2}\propto \Delta x^{-2/3}$. Our study indicates that the growth rate $\gamma$ follows the predicted theoretical scaling (see Fig. \ref{fig:gamma}). These results can only be confirmed by using an explicit kinematic viscosity coefficient so that $\mathrm{Re}$ can be measured directly, which, however, is beyond the scope of this work. 
	
	At early times, the magnetic energy is still subdominant with respect to the kinetic energy content of the flow. Therefore, the Lorentz force does not affect the evolution of the convection, and we do not observe any systematic difference in the velocity field between the MHD and the hydrodynamic simulations. This is the kinematic stage of the dynamo, which lasts for several convective turnover timescales, depending on the resolution of the grid.

	\subsection{Nonlinear phase of the dynamo}\label{sec:nonlinear}
	
	The amplification of the magnetic field due to the action of the small-scale turbulent dynamo proceeds until the Lorentz force becomes strong enough to start having a feedback effect on the flow. Such a change in the evolution of the dynamo happens when the magnetic energy approaches equipartition with the kinetic energy content of the eddies on the small scales of turbulence. Strong, small-scale magnetic fields inhibit the development of shear instabilities that feed the turbulent cascade and drive the dynamo amplification. The stretching of the magnetic field lines happens now on larger scales, where turbulence has not been quenched. Work done against the magnetic tension force by the turbulent convective flows sustains the magnetic field against numerical (resistive) dissipation, and the dynamo reaches saturation. By $t/\tau_\mathrm{conv}\,{\approx}\,15$, all of the MHD simulations presented here have entered this phase. This stage of the dynamo, however, does not represent a statistical steady state solution of the simulated setup. In fact, the continuous injection of entropy into the system by the heat source and the mixing processes that take place at the convective boundary (see Sect.~\ref{sec:me}) both contribute to the entrainment of material from the overlying stable layer. Therefore, the size,  mass, and  entropy content of the convective layer keep increasing over time. 

\begin{figure}
		\includegraphics[width=0.45\textwidth]{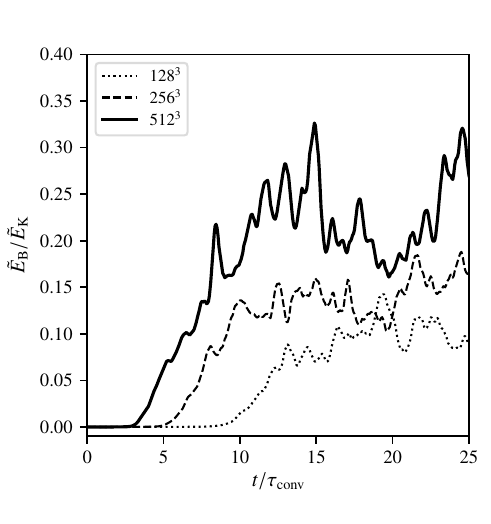}
		\caption{Time evolution of the mean magnetic-to-kinetic energy ratio inside the convective shell.}
		\label{fig:t-eb_ek}
	\end{figure}

\begin{figure}
		\includegraphics[width=0.45\textwidth]{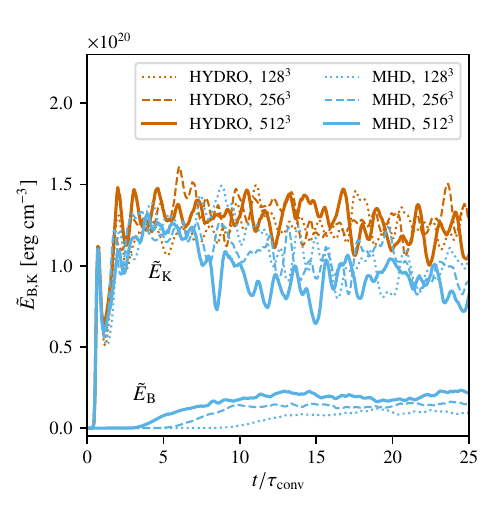}
		\caption{Time evolution of the mean kinetic energy density inside the convection zone, $\tilde{E}_\mathrm{K}$, for the purely hydrodynamic (vermilion) and MHD (light blue) simulations. The time evolution of the mean magnetic energy density in the MHD simulations, $\tilde{E}_\mathrm{B}$, is also shown.}
		\label{fig:t-eb-ek}
	\end{figure}

	The magnetic-to-kinetic energy ratio, shown in Fig.~\ref{fig:t-eb_ek}, reaches saturation with a mean value of ${\approx}\,0.22$ on the finest grid, with sporadic, intermittent episodes in which it reaches values as high as $0.33$. During the nonlinear phase of the dynamo, the mean kinetic energy density inside the convective shell in the MHD simulations,
	\begin{equation}
		\tilde{E}_\mathrm{K} = \frac{1}{2}\langle \rho |\bm{V}|^2 \rangle_\mathrm{conv},
	\end{equation}
    is on average $25\%$ lower than that in the hydrodynamic case on the 
    $512^3$ grid (see Fig.~\ref{fig:t-eb-ek}). We obtained this result by first computing the time average $\{\cdot\}$ of $\tilde{E}_\mathrm{K}$ over $t\,{\in}\,(15\tau_\mathrm{conv},25\tau_\mathrm{conv})$, and then by calculating 
    \begin{equation}
        \epsilon = \{\tilde{E}_\mathrm{K,\mathrm{MHD}}\}/\{\tilde{E}_\mathrm{K,\mathrm{HYDRO}}\}-1.
    \end{equation}
    However, the large temporal fluctuations that characterize $\tilde{E}_\mathrm{K}$ made it necessary to provide an error estimate on $\epsilon$ in order to prove the statistical significance of this result. The error on $\epsilon$ is a combination of the statistical uncertainties on $\{\tilde{E}_\mathrm{K,\mathrm{HYDRO}}\}$ and $\{\tilde{E}_\mathrm{K,\mathrm{MHD}}\}$, which we computed as follows. First we obtained the standard deviation $\sigma_\mathrm{K}$ of $\tilde{E}_\mathrm{K}$ over the selected time series for both the hydrodynamic and MHD setups. Second, we estimated the statistical uncertainty on the mean quantity  $\{\tilde{E}_\mathrm{K}\}$ by taking into account possible temporal correlations introduced by the turbulent nature of the convective flows. According to Fig. 4 of \cite{andrassy2022}, the autocorrelation function of the convective velocity drops to zero after a time shift  $\Delta t\,{\approx}\, \tau_\mathrm{conv}$, which suggests that there is approximately one independent realization of the convective flows per convective turnover. Thus, the uncertainty associated to  $\{\tilde{E}_\mathrm{K}\}$, $\tilde{\sigma}_\mathrm{K}$, was approximated as
    \begin{equation}
        \tilde{\sigma}_\mathrm{K} = \frac{\sigma_\mathrm{K}}{\sqrt{N_\mathrm{to}}},
    \end{equation}
    where $N_\mathrm{to}\,{=}\,10$ is the number of convective turnovers (see also Table~\ref{tab:ek}). Finally, we computed the variance of $\epsilon$,
    \begin{equation}
        \sigma^2_\epsilon    =   \Bigg(\frac{ \partial \epsilon }{ \partial \{\tilde{E}_\mathrm{K,\mathrm{HYDRO}}\}}\Bigg)^2\tilde{\sigma}^2_\mathrm{K,HYDRO}   +  \Bigg(\frac{ \partial \epsilon }{ \partial \{\tilde{E}_\mathrm{K,\mathrm{MHD}}\}}\Bigg)^2\tilde{\sigma}^2_\mathrm{K,MHD} .
    \end{equation}
    We find $\sigma_\epsilon \,{\approx}\, 3\%$, making a mean relative deviation of $25\%$ statistically significant (i.e., different from zero) by more than $8\sigma_\epsilon$.

    In the ILES approach, increasing the grid resolution reduces the amount of numerical resistivity introduced into the system, making the turbulent dynamo progressively more efficient. Consequently, the mean magnetic energy density in our simulations increases by a factor of two from the $128^3$ to the $512^3$ grid, where the typical strength of the magnetic field is  $\,{\approx}\,5\,{\times}\,10^{9}\ \mathrm{G}$ (see Table~\ref{tab:bfield}). On the other hand, $\tilde{E}_\mathrm{K}$ does not seem to show a significant resolution dependence when averaged over the saturated phase of the dynamo, $t\,{\in}\,(15\tau_\mathrm{conv},25\tau_\mathrm{conv})$, as can be noted from the values provided in Table~\ref{tab:ek}. Averaging over a wider time window could potentially reveal a statistically significant trend in  $\{\tilde{E}_\mathrm{K}\}$ with increasing spatial resolution. However, this is not possible with our current setup, in which the convective boundary approaches the upper domain boundary at late times, potentially altering the dynamics of the mixing region and producing unreliable results (see also Sect.~\ref{sec:Setup}).  We also note that a fixed time-averaging window, in principle, samples different evolutionary times of the dynamo depending on grid resolution. In fact, the oxygen shell does not have a statistical steady state solution, and the time at which the dynamo enters its nonlinear regime is resolution dependent. However, as visible in Fig.~\ref{fig:t-eb-ek}, neither $\tilde{E}_\mathrm{K}$ nor $\tilde{E}_\mathrm{B}$ show clear, long-term trends in the saturated phase of the dynamo. This result suggests that the secular evolution of the mean stratification can be neglected for this analysis since it does not seem to affect basic properties of the flows and of the dynamo.

\begin{figure*}
		\centering
		\includegraphics[width=0.98\textwidth]{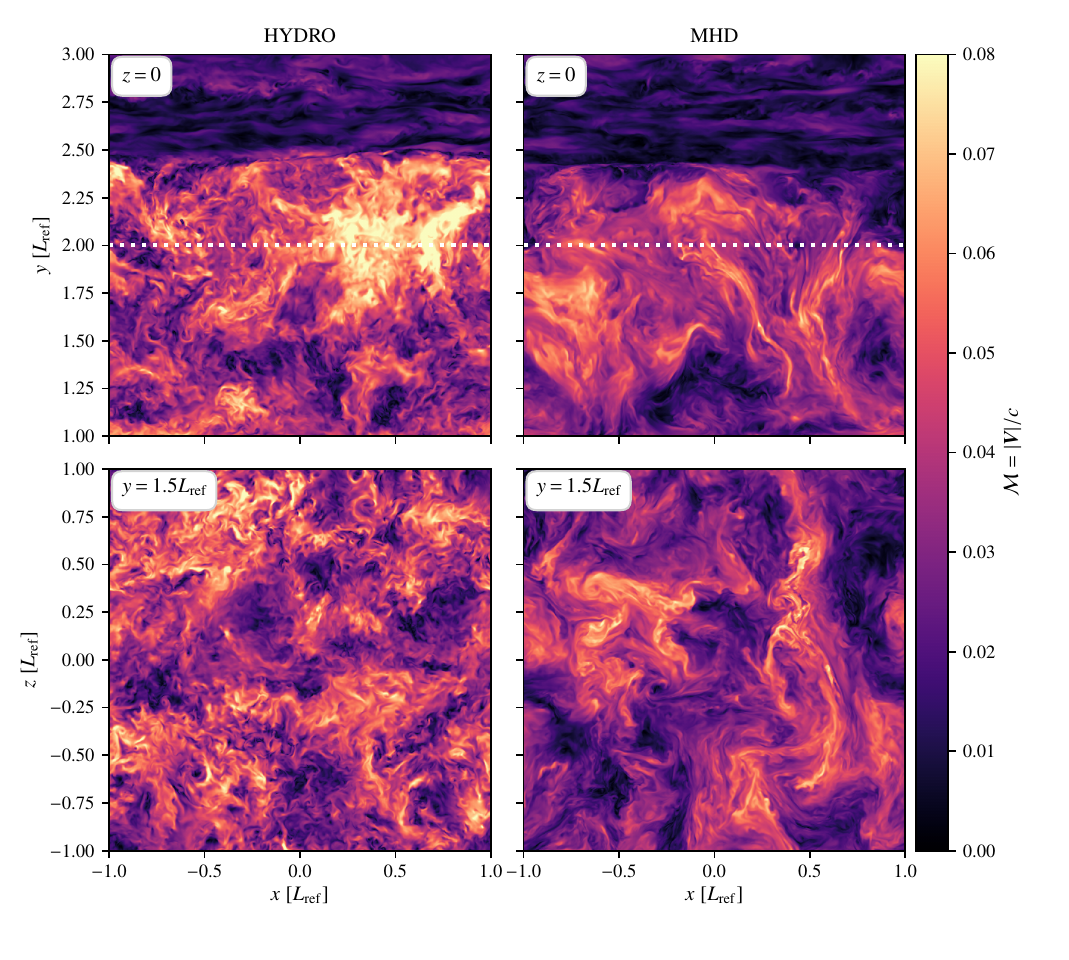}
		\caption{Snapshots of the Mach number on the $512^3$ grid at $t/\tau_\mathrm{conv}\,{=}\,22$. The panels on the left show results from the purely hydrodynamic simulation, while the panels on the right show the MHD simulation. The upper row shows vertical cuts taken in the $z\,{=}\,0$ plane, while the lower row shows a horizontal cut through the $y\,{=}\,1.5L_\mathrm{ref}$ plane. The white-dotted line indicates the initial position of the convective boundary.}
		\label{fig:mach-mhd-hydro}
	\end{figure*}

    The suppression of small-scale shear instabilities in the convective flows caused by the generated strong magnetic fields can be noted in Fig.~\ref{fig:mach-mhd-hydro}, where we compare snapshots of the Mach number taken from an MHD and a purely hydrodynamic simulation in the nonlinear phase of the dynamo. In contrast to the hydrodynamic case, where turbulence is essentially isotropic on spatial scales smaller than $L_\mathrm{ref}$, the velocity field in the MHD simulation is characterized by the presence of anisotropic, thread-like structures that extend over a large part of the convective shell.

	\begin{table}
		\caption{Mean kinetic energy density (in units of $\mathrm{erg}\ \mathrm{cm}^{-3}$) inside the convection zone, averaged over $t \in (15\tau_\mathrm{conv},25\tau_\mathrm{conv})$ for the indicated grid resolutions in the hydrodynamic and MHD cases.}
		\label{tab:ek}
		\centering
		\begin{tabular}{lll}
			\toprule
			$N$ & $\tilde{E}_\mathrm{K}/10^{20}\ (\mathrm{HYDRO})$ & $\tilde{E}_\mathrm{K}/10^{20}\ (\mathrm{MHD})$ \\
			\midrule
			$128$ & $1.25\pm0.03$ & $0.97\pm0.03$  \\
			$256$ & $1.26\pm0.02$ & $1.00\pm0.03$ \\
			$512$ & $1.22\pm0.03$ & $0.91\pm0.03$ \\
			\bottomrule
		\end{tabular}
		\vspace{0.5em}
		\tablefoot{The errors represent one standard deviation over the time series divided by $\sqrt{N_\mathrm{to}}$, where $N_\mathrm{to}\,{=}\,10$ is the estimated number of independent data points (one per convective turnover).}
	\end{table}

	\begin{table}
		\caption{Mean magnetic field strength inside the convection zone, averaged over $t \in (15\tau_\mathrm{conv},25\tau_\mathrm{conv})$ for the indicated grid resolutions.}
		\label{tab:bfield}
		\centering
		\begin{tabular}{ll}
			\toprule
			$N$ & $\langle|\bm{b}|/\sqrt{4\pi}\rangle_\mathrm{conv}\ [10^{9}\mathrm{G}]$  \\
			\midrule
			$128$ & $3.49\pm0.06$  \\
			$256$ & $4.24\pm0.06$  \\
			$512$ & $5.06\pm0.08$  \\
			\bottomrule
		\end{tabular}
		\vspace{0.5em}
		\tablefoot{The errors represent one standard deviation over the time series divided by $\sqrt{N_\mathrm{to}}$, where $N_\mathrm{to}\,{=}\,10$ is the estimated number of independent data points (one per convective turnover).}
	\end{table}

	\begin{figure}
		\centering
		\includegraphics[width=0.45\textwidth]{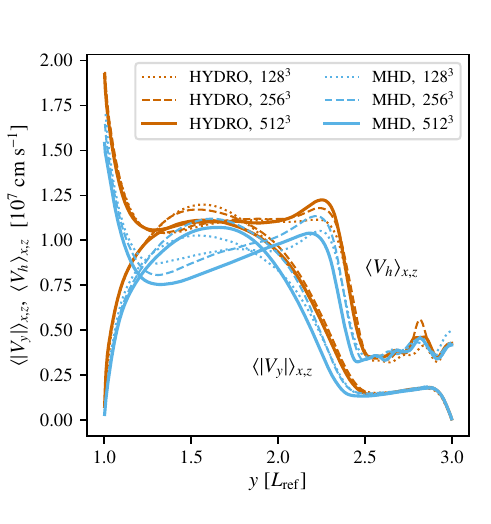}
		\caption{Vertical profiles of the horizontal ($V_h=\sqrt{V_x^2+V_z^2}$) and vertical ($V_y$) velocity components  averaged over $t \in (15\tau_\mathrm{conv},25 \tau_\mathrm{conv})$. Here, light blue is used to indicate the quantities extracted from the MHD simulations, whereas vermilion is used for the hydrodynamic simulations. An estimate of the statistical uncertainty associated to the averaged profiles can be inferred by looking at the typical dispersion among the hydrodynamic simulations, which represent a set of three independent (and numerically converged) realizations of the oxygen shell.}
		\label{fig:rad-vh-vy}
	\end{figure}

\begin{figure}
		\centering
		\includegraphics[width=0.45\textwidth]{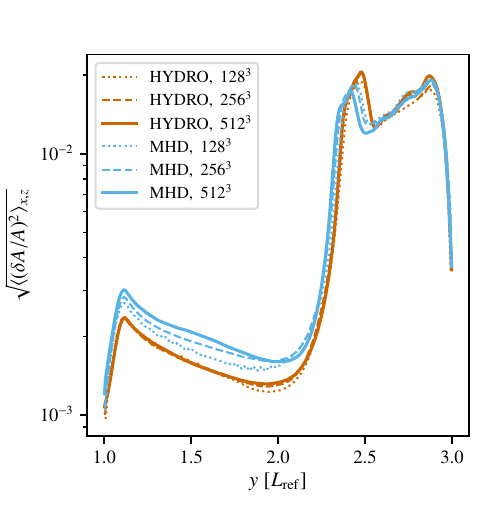}
		\caption{Vertical profiles of the root-mean-square relative pseudo-entropy fluctuation in the MHD (light blue) and purely hydrodynamic simulations (vermilion), averaged over $t \in (15\tau_\mathrm{conv},25 \tau_\mathrm{conv})$. The upper convective boundary is, on average, at the position of the peaks visible at \mbox{$y=(2.3\,{-}\,2.4)L_\mathrm{ref}$}.}
		\label{fig:rad-sfluxes}
	\end{figure}

\begin{figure}
		\centering
		\includegraphics[width=0.45\textwidth]{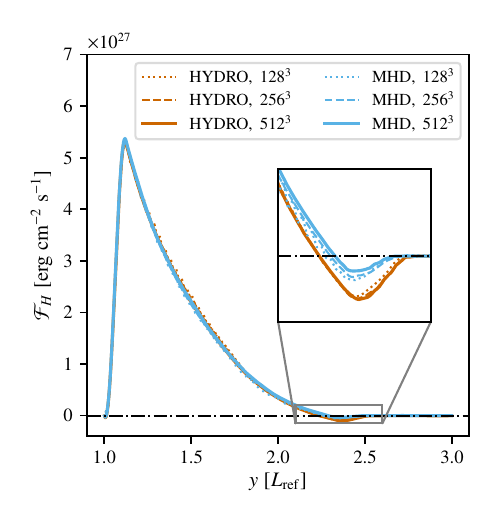}
		\caption{Vertical profiles of the vertical enthalpy flux ($\mathcal{F}_H$) averaged over $t \in (15\tau_\mathrm{conv},25 \tau_\mathrm{conv})$. Here, light blue is used to indicate the quantities extracted from the MHD simulations, whereas vermilion is used for the hydrodynamic simulations.}
		\label{fig:rad-fh}
	\end{figure}

	A comparison of horizontally averaged velocity profiles (see Fig.~\ref{fig:rad-vh-vy}) shows that horizontal velocities in the MHD case are reduced by as much as $30\%$ in the convective layer, as compared to the simulations without magnetic fields. Vertical velocities, instead, are diminished on average only by $10\%$. As visible in Fig.~\ref{fig:rad-sfluxes}, the partial suppression of the horizontal mixing in the MHD runs increases the magnitude of the root-mean-square entropy fluctuation inside the convection zone with respect to the hydrodynamic simulations. A larger contrast in the thermal content between up- and downflows in turn increases the efficiency of the convective energy transport. For this reason, despite the mild suppression of vertical velocities in the convection zone caused by the action of the dynamo, we do not observe any significant difference in the vertical enthalpy fluxes,
    \begin{equation}
        \mathcal{F}_H = \langle (e_\mathrm{int}+p) V_y \rangle_{x,z},
    \end{equation}
    between the MHD and the hydrodynamic setups (see Fig.~\ref{fig:rad-fh}). Only in the overshoot layer, where $\mathcal{F}_H\,{<}\,0$, the MHD simulations are characterized by a smaller unsigned enthalpy flux than their hydrodynamic counterpart. This result may be due to the combined effects of reduced flow speeds and entropy fluctuations at the upper convective boundary in the MHD case, as visible in Fig.~\ref{fig:rad-vh-vy} and Fig.~\ref{fig:rad-sfluxes}, respectively.

	\begin{figure}
	\centering
	\includegraphics[width=0.45\textwidth]{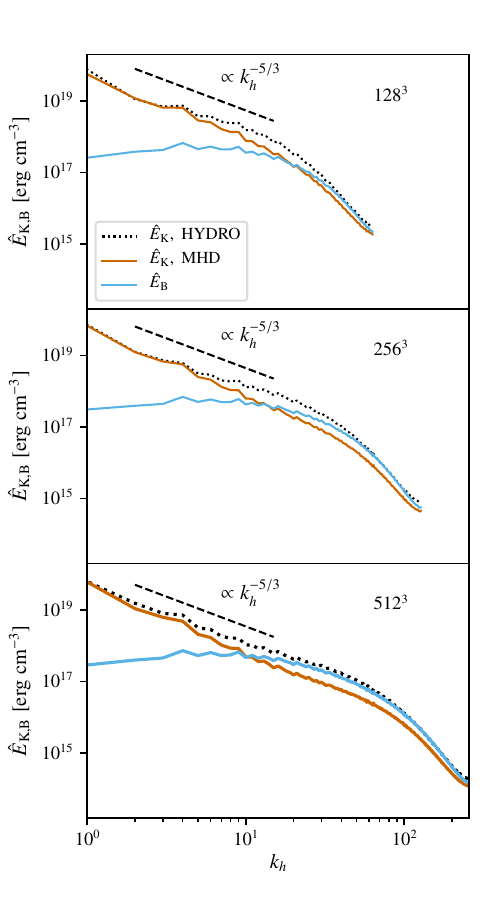}
	\caption{Magnetic (light blue) and kinetic (black-dotted for the hydrodynamic simulations, and vermilion for the MHD simulations) energy spectra computed in the $y\,{=}\,1.5L_\mathrm{ref}$ plane and averaged over the time interval $t \in (15\tau_\mathrm{conv},25 \tau_\mathrm{conv})$. Each panel shows the curves extracted from a different grid. The black-dashed lines indicate the Kolmogorov scaling law, $k_h^{-5/3}$.}
	\label{fig:spectra}
\end{figure}

\begin{figure*}
		\centering
		\includegraphics[width=\textwidth]{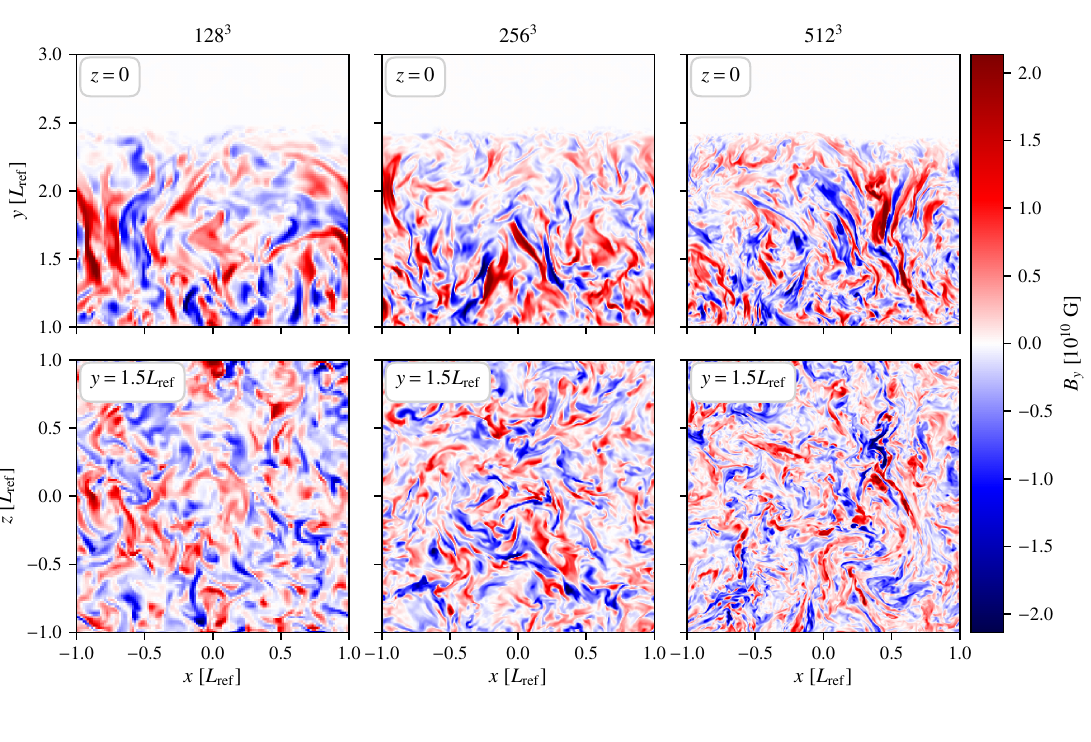}
		\caption{Distribution of $B_y$ on the $z\,{=}\,0$ plane (upper row) and the $y\,{=}\,1.5L_\mathrm{ref}$ plane (lower row) at $t/\tau_\mathrm{conv}\,{=}\,22$ for the indicated grid resolutions.}
		\label{fig:by-res}
	\end{figure*}

	In Fig.~\ref{fig:spectra}, we show the kinetic and magnetic energy spectra computed in the $y\,{=}\,1.5L_\mathrm{ref}$ plane, averaged over the saturated phase of the dynamo. The kinetic energy spectra resulting from the hydrodynamic simulations converge to the Kolmogorov scaling \citep[$k_h^{-5/3}$, ][]{kolmogorov1941} in the inertial range. The scale at which the power spectrum deviates from the Kolmogorov law due to the action of numerical dissipation becomes smaller as the resolution is progressively increased, as expected in the ILES approach. The kinetic energy spectra in the MHD simulations, instead, deviate from the hydrodynamic curves already at wavenumbers of $k_h\,{\approx}\,10$, where the dynamo is most efficient at converting kinetic energy into magnetic energy (i.e., it achieves maximum $|T_\mathrm{BKT}|/\hat{E}_\mathrm{K}$). The observed drop of kinetic energy in the MHD case corresponds to an increase of magnetic power in the inertial range, and the sum of the two energy contributions approximately resembles the kinetic energy spectrum in the hydrodynamic simulations. The wavenumber at which the magnetic-to-kinetic energy ratio becomes greater than unity decreases on finer grids\footnote{Energy spectra extracted from different planes in the convective shell show qualitatively similar results.}. On the grid with $512^3$ cells, the break-even point is at $k_h\,{=}\,10$, which corresponds to approximately half of the pressure scale height at $y\,{=}\,1.5L_\mathrm{ref}$. Unlike the case of forced, isotropic MHD turbulence \citep{haugen2004,schekochihin2004,iskakov2007,brandenburg2011}, our simulations of stratified convection also show a substantial amount (up to $30\%$) of magnetic power stored at wavenumbers $k_h\,{<}\,10$. This field component is generated by coherent structures in the form of large-scale up- and downflows that stretch the magnetic field lines over a large fraction of the size of the convection zone. The presence of a large-scale field component can be observed in Fig.~\ref{fig:by-res}, where we show vertical and horizontal cuts in $B_y$ for all our three grids. A small-scale component with mixed polarity also becomes more noticeable on progressively finer grids with reduced numerical dissipation.
	
	Both the horizontal and the vertical component of the magnetic field become stronger as the grid is refined (see Fig.~\ref{fig:rad-by-bh}). The magnetic field smoothly turns horizontal across the upper convective boundary, where the convective flows overturn due to the negative buoyant acceleration. At the bottom of the shell, the magnetic field is forced to be horizontal in order to retain its solenoidal property given the imposed boundary conditions. We note that the reflecting boundary forces the convective flows to abruptly change direction over a few computational cells, which artificially enhances the stretching and the compression of the magnetic field lines. This process, however, only affects the generation of magnetic energy in a narrow region close to the bottom boundary of the convective shell. In simulations of SSD action in the solar convection zone, including part of the underlying stable layer was shown to have little effect on the generation of the magnetic field as compared to simulations with closed boundary conditions \citep{hotta2017}. This author found that the imposed steep, positive entropy gradient across the solar overshoot region prevented convective plumes from penetrating into the stable layer deeper than a small fraction of the pressure scale height, at least on  the characteristic timescales set by convection. Thus, in those simulations, the bottom boundary of the solar convective region acted like a reflecting wall for the turbulent flows and the magnetic fields. Oxygen-burning shells of massive stars are also characterized by steep, stabilizing entropy gradients at their bottom boundary \citep{meakin2007,jones2017,varma2020}. In the model of \cite{jones2017} (which this setup is based on), the square of the Brunt-V\"ais\"al\"a frequency at the silicon-oxygen boundary was several times larger than that at the upper boundary of the oxygen shell \citep[see Fig. 4 of][]{jones2017}. Because this quantity is directly related to the buoyancy jump, entrainment of material from the underlying stable layer into the convection zone can easily be neglected over the timescales simulated here. All these considerations give us confidence that the chosen boundary conditions are well suited for the simulations presented in this study.

	The strength of the magnetic field is not uniform across the convective shell. As previously discussed, the magnetic energy approaches equipartition with the kinetic energy content of the turbulent eddies in the inertial range. Because the mean flow speed does not vary considerably across the convective shell,  the spatial dependence of the magnetic field strength is mostly set by the density stratification. Indeed, the vertical profile of the root-mean-square magnetic field rescaled by its local equipartition value ($B_\mathrm{eq}\,{=}\,\sqrt{\rho}V_\mathrm{rms}$) shows much less  dependence on $y$ as compared with the results shown in Fig.~\ref{fig:rad-eq}. The height-dependence of the dynamo action can also be seen in Fig.~\ref{fig:works}, where we show horizontal averages of the Lorentz work,
	\begin{equation}
		W_\mathrm{L} = \langle\bm{V}\cdot\big[(\nabla \times \bm{B})\times \bm{B}\big]\rangle_{x,z}.
	\end{equation}
	As a reference, we also show the buoyancy work,
	\begin{equation}
		W_\mathrm{b} = \langle V_y\, g_y \,\delta \rho \rangle_{x,z},
	\end{equation}
	where $\delta \rho$ is the density fluctuation. Buoyancy generates kinetic energy in the whole convective shell except in the overshoot layer, where it is responsible for the deceleration of the convective flows. On all grids, the magnitude of the Lorentz work 
	is maximum at the bottom of the convective shell and progressively drops to zero toward the upper convective boundary. $W_\mathrm{L}$ is negative throughout the whole convective layer, meaning that, on average, kinetic energy is everywhere converted into magnetic energy. Moreover, profiles of the Lorentz work approach convergence on the finest grids. This result confirms that most of the conversion of kinetic energy into magnetic energy happens on relatively large spatial scales, which are well resolved even with moderate grid resolutions.
	
	We note that the dynamo does not operate in the subadiabatic layer, although a seed field is present there as well. The turbulent structures created by the nonlinear breaking of IGWs (visible in Fig.~\ref{fig:mach-mhd-hydro}), which is one of the mechanisms that can excite an SSD in stable stratifications \citep{skoutnev2021}, are not
	efficient enough to build a significant magnetic field in these simulations. The magnetic field in the stable layer reaches saturation with average strengths of only two to ten times that of the initial seed field.

	\subsection{Impact of magnetic fields on the growth of the convective shell}
	\label{sec:me}
	
	The turbulent convective flows generated in this setup give rise to a rich variety of hydrodynamic processes at the upper convective boundary, including shear instabilities, breaking of surface gravity waves, and convective overshooting. These processes contribute to the entrainment of high-entropy material from the overlying stable layer into the convection zone, which causes the convective shell to grow in time (see also Fig.~\ref{fig:mach-mhd-hydro}). We computed the mass entrained per unit surface area inside the convection zone by using horizontal averages of the density and the passive tracer $\psi$ as in \cite{andrassy2022},
	\begin{equation}
		M_\mathrm{e}(t) = \int_{L_\mathrm{ref}}^{y_\mathrm{cb}(t)}  \bar{\rho}(y,t)\bar{\psi}(y,t) \mathrm{d}y.
	\end{equation}
	At each time, we assumed that the vertical coordinate of the upper convective boundary, $y_\mathrm{cb}$, was the position of the steepest gradient in $\bar{\psi}$. In Fig.~\ref{fig:me} we show the time evolution of the entrained mass for all the simulations run in this study. In the hydrodynamic case, numerical convergence  is  reached already on the lowest-resolved grid (with $128^3$ cells) within a maximum relative statistical uncertainty of $5\%$, which is consistent with the results obtained by \cite{andrassy2022}. Instead, the mass entrained in the MHD runs slightly decreases with increasing the grid resolution. A significant deviation between the MHD and the hydrodynamic simulations is visible only after $15 \tau_\mathrm{conv}$, when the dynamo has fully entered its nonlinear phase.
	By the time the simulations have finished, 
	the best resolved MHD setup has entrained $12\%$ less mass than the hydrodynamic runs, and the mass entrainment rate $\dot{M}_\mathrm{e}$ has been reduced by ${\approx}\,20\%$.

\begin{figure}
		\centering
		\includegraphics[width=0.45\textwidth]{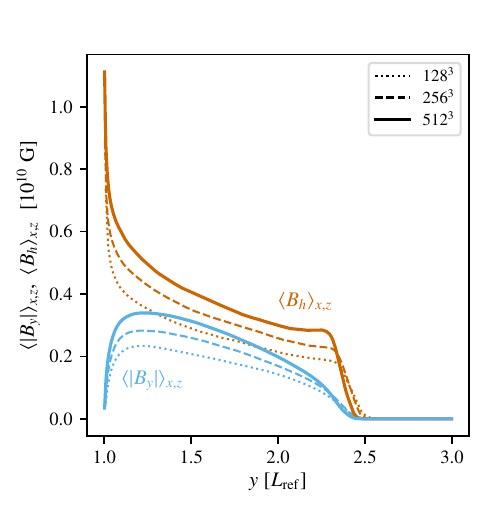}
		\caption{Vertical profiles of the horizontal ($B_h=\sqrt{B_x^2+B_z^2}$) and vertical ($B_y$)  magnetic field, averaged over the time interval $t \in (15\tau_\mathrm{conv},25 \tau_\mathrm{conv})$.}
		\label{fig:rad-by-bh}
	\end{figure}

\begin{figure}
		\centering
		\includegraphics[width=0.45\textwidth]{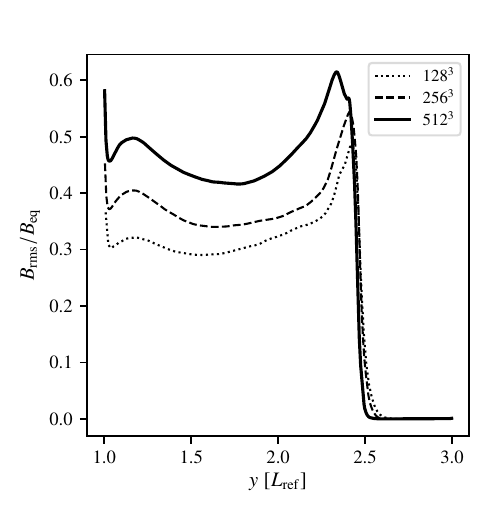}
		\caption{Vertical profiles of the root-mean-square magnetic field divided by the equipartition field, averaged over the time interval $t \in (15\tau_\mathrm{conv},25 \tau_\mathrm{conv})$.}
		\label{fig:rad-eq}
	\end{figure}

\begin{figure}
		\centering
		\includegraphics[width=0.45\textwidth]{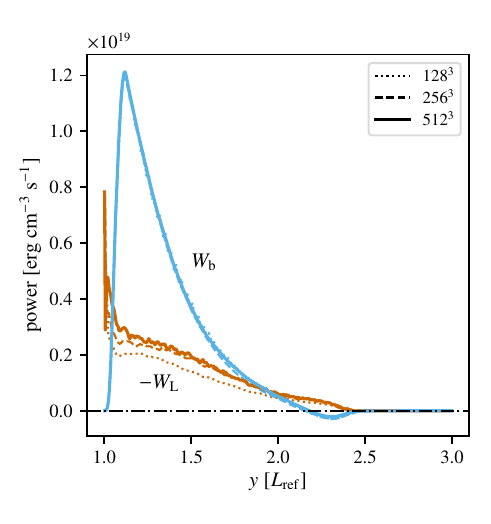}
		\caption{Vertical profiles of the buoyancy (light blue) and Lorentz work (vermilion) averaged over $t \in (15\tau_\mathrm{conv},25 \tau_\mathrm{conv})$ for the indicated grid resolutions.}
		\label{fig:works}
	\end{figure}

\begin{figure}
		\includegraphics[width=0.45\textwidth]{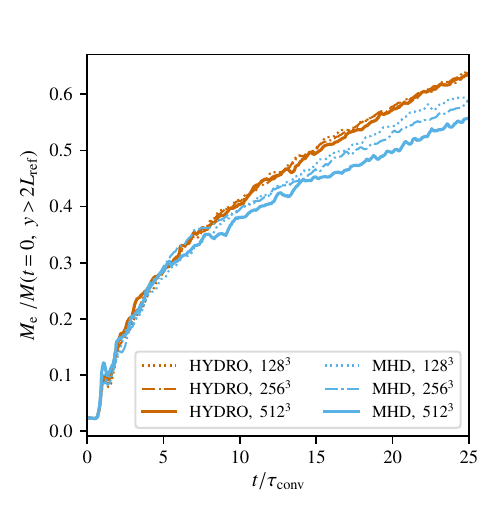}
		\caption{Time evolution of the mass entrained from the stable layer into the convection zone, rescaled by the total mass contained in the stable layer at $t\,{=}\,0\ \mathrm{s}$. Light blue is used for the MHD simulations, whereas vermilion is used for the purely hydrodynamic runs.}
		\label{fig:me}
	\end{figure}

	Because MHD effects do not dramatically reduce the mass entrainment rate at the upper convective boundary, finding the mechanisms responsible for the observed discrepancy in $M_\mathrm{e}$ between the MHD and the hydrodynamic results is challenging. One possible explanation is that convective flows in the MHD simulations have some of their kinetic energy converted into magnetic energy by the action of the SSD (see the discussion in Sect.~\ref{sec:nonlinear}), which reduces the amount of energy available to overcome the buoyancy of the entrained high-entropy material as compared to the hydrodynamic case \citep{spruit2015}. Furthermore, strong horizontal magnetic fields (see Fig.~\ref{fig:rad-by-bh}) could considerably reduce the growth rate of the shear instabilities that take place at the convective boundary, which are in part responsible for the mixing. Magnetic fields aligned with the shear flows, in fact, have a stabilizing effect against the growth of Kelvin--Helmholtz instabilities, especially if the Alfv\'enic Mach number $\mathcal{M}_\mathrm{Alf}\,{=}\,\sqrt{\rho}|\bm{V}|/|\bm{B}|$ is close to unity \citep{chandrasekhar1961,frank1996}. As shown in Fig.~\ref{fig:rad-eq}, the mean magnetic field at the upper convective boundary reaches values as high as $60\%$ of the equipartition field  ($\mathcal{M}_\mathrm{Alf}\,{\gtrsim}\,1.5$), so short-wavelength Kelvin--Helmholtz instabilities are likely to be partly suppressed. 
	
	As pointed out by a number of authors \citep{meakin2007,andrassy2020,horst2020a,andrassy2023}, some degree of mixing at stellar convective boundaries can be induced by nonadiabatic effects. In this setup, the mean entropy inside the convection zone increases by the action of the heat source, and eventually overcomes the entropy level of a narrow subadiabatic layer right above the upper convective boundary. This layer becomes negatively buoyant, so it sinks and gets mixed into the convection zone. This process enlarges the size of the convective shell over time as long as the source of entropy generation is active. In the work of \cite{andrassy2022}, it was estimated that by the end of the simulations, ${\approx}\,60\%$ of $\dot{M}_\mathrm{e}$ in this setup was due to the heating. This process is expected to operate regardless of the properties of the convective flows, so MHD processes would only be able to affect the remaining $40\%$ of the entrainment rate. In the absence of heating-induced mixing, magnetic fields would then suppress as much as $50\%$ of the mass entrainment.

	\section{Summary and discussion}\label{sec:Conclusions}
	
	We have run 3D simulations of turbulent convection, dynamo amplification and convective boundary mixing in an idealized oxygen-burning shell of a $25\ \mathrm{M}_\odot$ star. In particular, we have searched for possible MHD effects on the boundary mixing and the properties of the convective flows by performing a comparison between an MHD and a purely hydrodynamic setup. The numerical results have been carefully analyzed by means of a convergence study, in which the grid resolution was progressively increased from $128^3$ to $512^3$ cells. 
	
	Random stretching of the magnetic field lines due to the turbulent motions in the convective shell excites small-scale dynamo (SSD) action on all of the considered grids. The dynamo instability amplifies the seed field by ${\approx}\, 4$ orders of magnitude in a few convective turnover timescales. The kinematic phase of the dynamo ends when the magnetic field becomes strong enough to affect the evolution of the flows on the small scales of turbulence. During the saturated stage, the work done by fluid motions against magnetic tension forces sustains the magnetic field against numerical resistive dissipation. The saturated mean magnetic-to-kinetic energy ratio reaches values in the $20\,{-}\,30\%$ range. The magnetic field strength in the oxygen shell moderately increases with the grid resolution, and it has characteristic values of ${\sim}\,10^{10}\ \mathrm{G}$ in the $512^3$ simulation. Such strong fields partly suppress the small-scale isotropic features in the velocity field typical of turbulent convection in hydrodynamic simulations. The resulting flows present anisotropic, thread-like structures that extend over a large fraction of the convective shell. The magnetic fields generated during the oxygen burning stage can be further amplified if parts of the oxygen shell end up collapsing onto a neutron star. By assuming a simple flux-freezing model, we estimate that the magnetic field strength at the surface of the neutron star would be on the order of $10^{15}-10^{16}$ G, which is in agreement with values inferred from observations of magnetars \citep{kouveliotou1998,woods2006,olausen2014}.
	
	Vertical and horizontal fluid velocities in the bulk of the convective layer in the MHD simulations are reduced, as compared to the hydrodynamic runs,  on average by $10\%$ and $20\%$, respectively. The fact that the dynamo does not have the same impact on the different fluid velocity components could be related to the transport of thermal energy inside the convection zone. In fact, in order for convection to transport the excess heat deposited by the energy source outward with partly suppressed vertical velocities, the thermal content of the buoyant flows must be enhanced. Because the heat source is unchanged in the MHD simulations, this can only be achieved by reducing the horizontal mixing of entropy between the up- and the downflows. Indeed, we observe that root-mean-square entropy fluctuations are systematically enhanced in the MHD simulations, which in turn increases the thermal contrast between the convective plumes. Consequently, we do not observe a significant impact of magnetic fields on the vertical enthalpy fluxes.
	
	Power spectra computed in the bulk of the convective shell reveal that $30\%$ of the total magnetic energy is stored at spatial wavenumbers $k_h\,{<}\,10$. Such a large-scale field component is generated by large-scale convective cells, which can efficiently stretch the magnetic field lines on length scales comparable to the size of the convection zone. The kinetic energy spectra in the MHD simulations deviate from the Kolmogorov law ($k_h^{-5/3}$) in the inertial range, where the efficiency of the dynamo is maximum. On the finest grid, the magnetic energy becomes greater than the kinetic energy at a spatial wavenumber of $k_h\,{\approx}\,10$, corresponding to half of a pressure scale height in the convection zone.
	
	The mass entrained into the convection zone in the MHD case is smaller by $12\%$ than that in the hydrodynamic setup. The partial suppression of the mixing at the convective boundary correlates with the average strength of the magnetic field in the convection zone. It is possible that the reduction of the kinetic energy of the convective flows caused by the action of the small-scale turbulent dynamo and the presence of magnetic fields with strengths up to $60\%$ of the equipartition value at the convective boundary contribute to the partial suppression of the mixing. By the end of the simulations, the mass entrainment rate is reduced by $20\%$ with respect to the hydrodynamic simulations.
	
	In our simulations, SSDs seem to have only a mild effect on the growth of the convective oxygen shell. This is consistent with the findings of \cite{varma2020}, who ran global MHD simulations of an oxygen shell in an $18\ \mathrm{M}_\odot$ star. Overall, these authors did not observe any significant impact of the small-scale turbulent dynamo on the properties of the convective flows as compared to a non-MHD simulation. Those results, however, may have been affected by low effective Reynolds numbers, which are typical for global simulations of turbulent convection at moderate grid resolutions. Our ``box-in-a-star'' approach, used in combination with special numerical solvers optimized for tackling stratified, subsonic magnetoconvection, allows us to achieve much higher effective resolution than that obtained by \cite{varma2020}. We find that the small-scale turbulent dynamo reduces the kinetic energy content of the convective shell  by $25\%$ on average and significantly changes the topology of the velocity field with respect to the purely hydrodynamic problem. These results are particularly important in the context of the ``perturbation-aided'' explosion mechanism, whose efficiency is set by both the magnitude of the convective velocities and the typical spatial scales of convection in the burning shells of the supernova progenitor \citep{muller2017,couch2020}. 
	
	One point of concern is related to the usage of the Implicit-Large-Eddy-Simulation (ILES) approach in this study, which gives rise to effective magnetic Prandtl numbers $\mathrm{Pr}_\mathrm{m}=\nu/\eta$ close to or even larger than unity. Such large values of $\mathrm{Pr}_\mathrm{m}$ are an overestimation of the actual conditions found in oxygen-burning shells. Based on the analytic expressions of the viscous and resistive coefficients provided in \cite{augustson2019}, we estimate that realistic magnetic Prandtl numbers in the oxygen shell considered here range from $\mathrm{Pr}_\mathrm{m}\,{=}\,0.001$ to $\mathrm{Pr}_\mathrm{m}\,{=}\,0.1$. It is well known that certain properties of the small-scale turbulent dynamo are sensitive to the value of $\mathrm{Pr}_\mathrm{m}$. In particular, the strength of the saturated magnetic field is often found to be an increasing function of $\mathrm{Pr}_\mathrm{m}$ \citep{schekochihin2004,brandenburg2011,kapyla2018}. Unfortunately, no general consensus has been reached so far as to the behavior of SSDs at low $\mathrm{Pr}_\mathrm{m}$, which is likely setup dependent. More MHD simulations of the oxygen-burning phase with realistic Prandtl numbers are therefore required in order to properly establish the impact of magnetic fields on the dynamics of the convective shell. Finally, we did not include a nuclear-burning network in our simplified setup, so further investigation on possible indirect effects of magnetic fields on the nuclear energy generation and nucleosynthesis is certainly needed.

	\begin{acknowledgements}
		GL, RA, JH, and FKR acknowledge support by the Klaus Tschira
		Foundation. PVFE was supported by the U.S. Department of Energy
		through the Los Alamos National Laboratory (LANL). LANL is operated by
		Triad National Security, LLC, for the National Nuclear Security Administration
		of the U.S. Department of Energy (Contract No. 89233218CNA000001).
		This work has been assigned a document release number LA-23-27290. The authors would like to thank Hirschi R., Rizzuti F., and Varma V. (Keele University) for fruitful discussions and hosting a visit to Keele University.
	\end{acknowledgements}
	
	\bibliographystyle{aa}
	\bibliography{oxygen-shell-MHD}
	
	\begin{appendix}

	\end{appendix}

\end{document}